\newcommand{\diracslash}[1]{#1\llap{/\kern2pt}}

\newcommand{\be}{\begin{equation}}
\newcommand{\ee}{\end{equation}}
\newcommand{\bea}{\begin{align}}
\newcommand{\eea}{\end{align}}
\newcommand{\ba}[1]{\begin{array}{#1}}
\newcommand{\ea}{\end{array}}

\newcommand{\bt}{\begin{tabular}}
\newcommand{\et}{\end{tabular}}

\newcommand{\beas}{\begin{align*}}
\newcommand{\eeas}{\end{align*}}

\documentclass[preprint,prd,aps,floats,nofootinbib,floatfix]{revtex4}

\DeclareSymbolFont{rsfs}{U}{rsfs}{m}{n}
\DeclareSymbolFontAlphabet{\mathrsfs}{rsfs}

\usepackage{amsmath}
\usepackage{graphicx}
\usepackage{multirow}
\usepackage{graphicx,epstopdf}

\usepackage{graphicx}
\usepackage{float}

\usepackage[utf8]{inputenc}
\usepackage[english]{babel}
\usepackage{hyperref}
\hypersetup{
    colorlinks=true,
    linkcolor=blue,
    filecolor=magenta,     
    citecolor=blue, 
    urlcolor=blue,
}
 
\usepackage{cleveref}

\begin{document}

\title{Possibility of $\rho$ Meson Condensation in Neutron Star: Unified Approach of Chiral SU(3) Model and QCD Sum Rules}

 \author{Shivam}
\email{shivamchoudhary413@gmail.com}
\affiliation{Department of Physics, Dr. B R Ambedkar National Institute of Technology Jalandhar, 
 Jalandhar -- 144011,Punjab, India}
\author{Arvind Kumar}
\email{iitd.arvind@gmail.com, kumara@nitj.ac.in}
\affiliation{Department of Physics, Dr. B R Ambedkar National Institute of Technology Jalandhar, 
 Jalandhar -- 144011,Punjab, India}

\def\be{\begin{equation}}
\def\ee{\end{equation}}
\def\bearr{\begin{align}}
\def\eearr{\end{align}}
\def\zbf#1{{\bf {#1}}}
\def\bfm#1{\mbox{\boldmath $#1$}}
\def\hf{\frac{1}{2}}
\def\kp{\zbf k+\frac{\zbf q}{2}}
\def\km{-\zbf k+\frac{\zbf q}{2}}
\def\hwo{\hat\omega_1}
\def\hwt{\hat\omega_2}

\begin{abstract}
In the present work the conjunction of chiral SU(3) model with QCD sum rules is employed to explore the possibility of $\rho$ meson condensation in neutron stars. The quark and gluon condensates in terms of which the in-medium masses of $\rho$ mesons can be expressed are calculated using the chiral SU(3) model in the charge neutral matter which is relevant for neutron stars. It is observed that condition of $\rho$ meson condensation is satisfied for the density of about 7$\rho_{0}$, where $\rho_{0}$ is the nuclear saturation density. In the end, a brief qualitative discussion of the magnetic field is also involved to check out for the further possibility of $\rho$ meson condensation. 

\end{abstract}

\maketitle

\maketitle

\section{Introduction}
\label{intro}
The study of astrophysical compact objects like neutron stars plays a very important role in understanding the strong interaction physics at high baryonic densities. The heavy ion collision experiments, for example,  Large Hadron Collider (LHC) and Relativistic Heavy Ion Collider (RHIC) aims to explore the Equation of State (EoS) of dense matter at high temperature and very low baryonic density whereas Compressed Baryonic Matter (CBM) experiment of FAIR project, Nuclotron-based Ion Collider Facility (NICA) at Dubna, Russia and Japan Proton Accelerator Research Complex (J-PARC) will focus on high baryonic density and relatively moderate temperature. The recent detection of gravitational waves by LIGO from neutron star merger puts more strong constraint on the EoS of dense medium \cite{Annala2018} and this has triggered much more theoretical calculations in the field \cite{Abbot2017,Pal,Benjamin,Kyu,Ho2018}. The data obtained on the masses and radii of the neutron star from the Neutron Star Interior Composition Explorer (NICER) experiment at international space station impose the significant constraint on the EoS  \cite{Gendreau, Baubock, Miller, Ozel, Grindlay, Watts}. 
%These constraints play a vital role in understanding the recently detected gravitational waves \cite{Baym}. 
In order to explain the observed data from various other laboratories like LIGO-Virgo \cite{BP, Abb, BPA}, LIGO-India \cite{Iyer}, GEO \cite{Hough} we need to mould our models and the existence of these high mass neutron stars will also hold out restriction on theoretical calculations \cite{Baym}. Furthermore, the importance of studying the neutron star like objects lies in the fact that at the core of these objects the density can be up to many times of the nuclear saturation density which
is not achieved in human-made laboratories and ultimately makes these objects a natural laboratory to understand the strong interaction physics \cite{Lattimer}. Whereas, the strongly interacting matter in heavy ion collisions exists at finite temperature and baryonic density, the matter in the neutron star is at high baryonic density and zero temperature in the ground state. 
\par 
Numerous theoretical non-perturbative approaches, for example,  
 Quark Meson Coupling (QMC) model  \cite{Yue,Guichon,HW,Panda,Tsushima,Sibirtsev},
 Polyakov Quark-Meson Coupling model \cite{Schaefer,Mohan}, 
 chiral effective models  \cite{Dexheimer,Mallik,Thorsson}, QCD sum rules, coupled channel approach \cite{Tol763,Tol635,Tol77,Hofmann}, NJL and PNJL models \cite{Nambu,Kukushima,Kashiwa,Ghosh} etc.,  
are employed to investigate the properties of hadrons in dense interacting environment.
The idea of condensate formation in dense matter was first proposed a long ago by Kaplan and Nelson \cite{Kaplan}. The inspiration for the meson condensation comes from strong attractive interaction of mesons with nucleons. Due to this attractive interaction, the effective energy of the mesons becomes low in the matter as compared to that in the vacuum. This notion provides a base for the kaon-antikaon condensation in dense matter \cite{Pal20}. The reason for the possibility of kaon condensation in a neutron star is its attractive interaction with nucleons both in vector and axial vector channels  \cite{Kaplan2}. With the increasing baryonic density of the medium, the in-medium energy of the antikaon decrease whereas the chemical potential of electron increases. Beyond a particular density at which both become equal, antikaon can play the role of charge neutrality and energetically antiakons will be favored over electrons.  This will lead to the phenomenon of antikaon condensation in neutron stars and will have the consequences on EoS and hence, the mass and radius of these objects \cite{Pal20,AK}.

 The in-medium mass reduction of light vector $\rho$ mesons predicted by various studies\cite{Lee,Hayacizaki,Asakawa} motivate to investigate the possibility of $\rho$ meson condensation in neutron star \cite{Mallick2015,EE}. 
The \cite{starc1} results on $\rho^{0} \rightarrow \pi^{+} \pi^{-}$  production   in the peripheral Au-Au collisions at $\sqrt{S} = 200$ GeV were reported and attractive mass-shift  of about $70$ MeV was observed.
 The enhancement in the production of low mass invariant dileptons in the heavy ion collision experiment like CERES at CERN \cite{Agakichiev} provides the possibility in the mass reduction of $\omega$ mesons but width broadening in the $\rho$ mesons \cite{Ko}. Future heavy-ion collision experiments may help to clarify the situation on the mass shift of $\rho$ mesons further.
%  One estimate of such enhanced dileptons formation in HIC is the potential reduction in the mass of vector mesons \cite{Mallick2015}.
%  At such high baryonic density, the spontaneously broken chiral symmetry \cite {Greiner, Kumar2010} is expected to be restored due to the presence of high quark condensates \cite{Mallick2015}. 
%  However, the exact observation for such a symmetry restoration is not well-defined  \cite {Mallick2015}. 
 In order to look for the possibility of $\rho$ meson condensation, an approach similar to the antikaons can be used where we try to anticipate a comparison between effective electron chemical potential and in-medium masses of $\rho$ mesons. The presence of these antikaon or $\rho$ meson condensates not only soften the EoS which will impose the constraint on the mass of star \cite{Schaffner,Glendenning} but will also affect the restoration of chiral symmetry \cite{Thorsson}. Thus, the possibility of $\rho$ meson condensation in the highly dense matter cannot be simply ignored.

  Hatsuda and Lee used QCD sum rules to calculate the in-medium masses of $\rho$ meson by assuming a linear density approximation and a considerable amount of decrease is observed in the in-medium masses of $\rho$ mesons \cite{Lee}. Consequently, different studies on using QCD sum rules for investigating the properties of vector mesons evolved\cite{Mallik,Stefan,KK}. In QCD sum rule analysis, in-medium  properties of different mesons are evaluated in terms of medium modification of quark and gluon condensates\cite{Weise}. Generally, these calculations employ linear density approximation and results are normally valid upto nuclear saturation density. In our present paper, we will use non-linear effective chiral SU(3) model\cite{Schramm2008,HANAUSKE2000} to calculate the density dependence of quark and gluon condensates. Within the model, we have 
   non-linear interaction terms of the scalar fields  $\sigma$, $\zeta$, $\delta$, $\chi$ and also for vector fields $\omega$ and $\rho$. The scalar iso-vector field $\delta$ and vector iso-vector field $\rho$ contribute in isospin asymmetric matter and since, in the charge neutral matter of neutron stars the density of proton and neutrons are different, these fields
   are important to consider in the model calculations. Also, as we will see in next section the dilaton field $\chi$ will be used to write the expressions for gluon condensates through the trace anomaly property of QCD.
   The scalar quark and gluon condensates calculated in charge neutral matter relevant for neutron stars using chiral model will be further used in the QCD sum rules to find the density dependence of the masses of vector $\rho$ mesons.
    %The models like linear $\sigma$-$\omega$ model in Hartree approximation \cite{Zhang}, Quark Meson Coupling (QMC) in relativistic mean field approximation \cite{Yue,Guichon,Hong,Panda,Tsushima,Sibirtsev}, chiral SU(3) model \cite{Dexheimer} etc. were used to study the strong interaction physics and models like \cite{Walecka, Bodmer, Boquta, Rufa, Ring, Serot, P, Particle, Weber,Savushkin} are suitable to use in the high density and finite temperature. These models are also expected to work fairly well to hadronic physics in neutron star. Furthermore, various other QCD properties dependent models are also adopted to study the hadronic matter at finite temperature and high baryonic density. A few of such are Walecka model \cite{Walecka},coupled-channel approach \cite{Tol763,Tol635,Tol77,Hofmann} and Polyakov Quark-Meson Coupling model \cite{Schaefer,Mohan}.
%   In the present investigation, we used the chiral SU(3) model in the relativistic mean field approximation to calculate the scalar and vector fields ($\sigma$, $\zeta$, $\delta$, $\chi$, $\omega$ and $\rho$) and QCD sum rules to obtain the in-medium masses of $\rho$ mesons. At first we will use the values of the fields to calculate the in-medium scalar quark and gluon condensates. These condensates are then be utilised in QCD sum rules to calculate the in-medium mass of $\rho$ meson.
    The in-medium mass of $\rho$ meson and electron chemical potential are then compared to check out for any possibility of $\rho$ meson condensation in a neutron star.
%     The decent interpretation of meson-meson interactions  \cite{NA} and in-medium properties of kaons in nuclear matter \cite{Ramos,Kaos} using chiral SU(3) model tells its success in explaining the strong hadronic interaction. In the case of multi hypernuclei objects, this model also provides an entirely different view on the properties of these objects  \cite{Greiner}. 
    The chiral SU(3) model has already been widely used by many authors to study the hadronic properties in nuclear and strange hadronic matter and also the properties of neutron stars \cite{Dexheimer, Sanyal, Schramm2008, HANAUSKE2000,Greiner}.
\par This paper is organised in the following way: In section \ref{sec:2}, we will discuss the formalism opted for the present investigation of calculating the in-medium masses of $\rho$ mesons and its condensation. The chiral SU(3) model in neutron star matter including the necessary conditions of charge neutrality and beta equilibrium is discussed in subsection \ref{subsec:2.1}. In subsection \ref{subsec:2.2}, we will discuss the QCD sum rules used to calculate the in-medium masses of $\rho$ mesons in the present work. The result obtained from the present investigation are discussed in section \ref{sec:3}.

\section{Formalism}
\label{sec:2}
\subsection{Chiral SU(3) Model in Neutron Star}
\label{subsec:2.1}
\quad Quantum Chromodynamics, the theory of strong interaction due to large coupling constant is not directly applicable in non-perturbative regime  \cite{Greiner,Papazoglou}. Therefore, effective models constrained by various properties of QCD are formulated to probe the physics of non-perturbative regime \cite{Balazs, Greiner,Dexheimer, Sanyal, Schramm2008, HANAUSKE2000}. Chiral SU(3) model is one such effective field theoretical model based on broken scale invariance and spontaneous chiral symmetry breaking  \cite{kumar2015}. The property of broken scale invariance of QCD is incorporated by adding a glueball field $\chi$ at tree level through logarithmic potential  \cite{Schechter1980, Greiner}. The Lagrangian density in chiral SU(3) model consists of various interaction terms among different field particles and is given by
\begin{align}
 \mathcal{L}=\mathcal{L}_{Kin} + \mathcal{L}_{i,Scal} + \mathcal{L}_{i,Vec} + \mathcal{L}_{Scal} + \mathcal{L}_{Vec} + \mathcal{L}_{SB}
 \label{eq1}
\end{align}
where the first term $\mathcal{L}_{Kin}$ in the above Lagrangian  represents the kinetic energy of nucleons and scalar ($\sigma$, $\zeta$, $\delta$ and $\chi$) and vector ($\omega$ and $\rho$) mesons, $\mathcal{L}_{i,Scal} + \mathcal{L}_{i,Vec}$ represents the interaction of nucleons with scalar and vector mesons, $\mathcal{L}_{scal}$ represents the meson-meson interaction which is responsible for the spontaneous breaking of chiral symmetry, $\mathcal{L}_{vec}$ represents the self interaction of vector mesons and is responsible for generating the mass of spin-1 mesons through the interaction with spin-0 mesons and the last term, $\mathcal{L}_{SB}$ simply explains the explicit symmetry breaking of U(1), SU(2) and chiral symmetry, and is responsible for the masses of pseudoscalar mesons. Explicitly, various terms are given by \cite{Dexheimer,Schramm2008}
\begin{align}
\mathcal{L}_{i,Vec} + \mathcal{L}_{i,Scal} =-\bar{\psi}_i[m_{i}^{*} + g_{\omega i}\gamma_{0}\omega + g_{\rho i}\gamma_{0}\tau_3 \rho]{\psi}_i,
\label{eq2}
\end{align}

\begin{align}
{\mathcal L} _{Scal}  = & -\frac{1}{2} k_{0}\chi^{2} \left( \sigma^{2} + \zeta^{2} 
+ \delta^{2} \right) + k_{1} \left( \sigma^{2} + \zeta^{2} + \delta^{2} 
\right)^{2} \nonumber\\
+& k_{2} \left( \frac {\sigma^{4}}{2} + \frac {\delta^{4}}{2} + 3 \sigma^{2} 
\delta^{2} + \zeta^{4} \right) 
+ k_{3}\chi\left( \sigma^{2} - \delta^{2} \right)\zeta \nonumber\\
-& k_{4} \chi^{4} 
 -  \frac {1}{4} \chi^{4} {\rm {ln}} 
\frac{\chi^{4}}{\chi_{0}^{4}}
+ \frac {d}{3} \chi^{4} {\rm {ln}} \Bigg (\bigg( \frac {\left( \sigma^{2} 
- \delta^{2}\right) \zeta }{\sigma_{0}^{2} \zeta_{0}} \bigg) 
\bigg (\frac {\chi}{\chi_0}\bigg)^3 \Bigg ),
\label{eq3}
\end{align}
\begin{align}
\mathcal{L}_{Vec}=\frac{1}{2}\Big(m_{\omega}^2 \omega^2 + m_{\rho}^2 \rho^2\Big)\Bigg(\frac{\chi}{\chi_{0}}\Bigg)^2 + g_{4}(\omega ^4 + 6\omega^2 \rho^2 + \rho^4),
\label{eq4}
\end{align}
%\begin{align}
%\mathcal{L}_{SB}=-\Bigg(\frac{\chi}{\chi_{0}}\Bigg)^2[m_{\pi}^2f_{\pi}\sigma+(\sqrt{2}m_{K}^2f_{K}-\frac{1}{\sqrt2}m_{\pi}^2f_{\pi})\zeta],
%\end{align}
\begin{align}
\mathcal{ L} _{SB}= -\left( \frac {\chi}{\chi_{0}}\right)^{2} 
\left( \frac{1}{2} m_{\pi}^{2} 
f_{\pi} \left( \sigma + \delta \right) +
\frac{1}{2} m_{\pi}^{2} 
f_{\pi} \left( \sigma - \delta \right)
 + \big( \sqrt {2} m_{K}^{2}f_{K} - \frac {1}{\sqrt {2}} 
m_{\pi}^{2} f_{\pi} \big) \zeta \right),
\label{exp_lterm}
\end{align}

where $m_{i}^{*}$ is the effective mass of $i^{th}$ nucleon. The values of various coupling constants, $i.e.$, $g_{fi}$ ($f$ represents the field and $i$ represents the nucleon) and other constants are chosen in such a way so as to reproduce the vacuum masses of hadrons, hyperon potentials, nuclear saturation properties like binding energy per nucleon ($B/A$), asymmetry energy \cite{Greiner,Papazoglou,Dexheimer}. Also, $m_{\pi}$, $f_{\pi}$ and $m_{K}$, $f_{K}$ are masses and decay constants of $\pi$ and $K$ meson, respectively. The values of these constants are listed in \cref{table1}.
The thermodynamical potential per unit volume in grand canonical ensemble for the medium of neutron stars is given by 
\begin{equation}
\frac{\Omega} {V}=-{\mathcal L}_{vec} - {\mathcal L}_0 - {\mathcal L}_{SB}-{\mathcal{V}}_{vac} - \beta_{i} - \alpha_{e}. 
\label{eq_grand1}
\end{equation}
In above, $\mathcal{V}_{vac}$ is the vacuum energy which is subtracted to obtain the zero vacuum energy, $\alpha_{e}$ and $\beta_{i}$ are the energy terms arising from the electrons and nucleons. The corresponding expressions are given as
\begin{align}
\beta_{i}=\sum_{i}\frac{\gamma_{i}}{48\pi^{2}}\Big[\Big(2k_{fi}^{3}-3m_{i}^{*2}k_{fi}\Big)\mu_{i}^{*}+3m_{i}^{*4}\ln\Big\{\frac{k_{fi}+\mu_{i}^{*}}{m_{i}^{*}}\Big\}\Big],
\label{eq5}
\end{align}
and
\begin{equation}
\alpha_{e}=\frac{\mu_{e}^{4}}{12\pi^{2}},
\label{eq6}
\end{equation}
where $k_{fi}$ represents the fermi momentum of the nucleon species and $\gamma_{i}$ is the fermionic spin-isospin degeneracy factor which is taken to be equal to 2. Furthermore,  $E_{fi}^{*}$=$\sqrt{k_{fi}^{2}+m_{i}^{*2}}$ and $\mu_{i}^{*}=\mu_{i}-g_{\omega i}\omega-g_{\rho i}\tau_{3}\rho$, are the single particle energy and effective chemical potential, respectively \cite{Schramm2008,HANAUSKE2000}. The equations of motion corresponding to scalar and vector fields can be obtained by minimizing the grand potential given by \cref{eq_grand1} because this will correspond to an equilibrium condition  \cite{Wang2003}. We have following equations for $\sigma, \zeta, \delta, \chi, \omega$ and $\rho$: 
\begin{align}
& k_{0}\chi^{2}\sigma-4k_{1}\left( \sigma^{2}+\zeta^{2}
+\delta^{2}\right)\sigma-2k_{2}\left( \sigma^{3}+3\sigma\delta^{2}\right)
-2k_{3}\chi\sigma\zeta \nonumber\\
-&\frac{d}{3} \chi^{4} \bigg (\frac{2\sigma}{\sigma^{2}-\delta^{2}}\bigg )
+\left( \frac{\chi}{\chi_{0}}\right) ^{2}m_{\pi}^{2}f_{\pi}
-\sum_{i} g_{\sigma i}\rho_{i}^{s} = 0,
\label{si_eq}
\end{align}
\begin{align}
&k_{0}\chi^{2}\zeta-4k_{1}\left( \sigma^{2}+\zeta^{2}+\delta^{2}\right)
\zeta-4k_{2}\zeta^{3}-k_{3}\chi\left( \sigma^{2}-\delta^{2}\right)\nonumber\\
-&\frac{d}{3}\frac{\chi^{4}}{\zeta}+\left(\frac{\chi}{\chi_{0}} \right)
^{2}\left[ \sqrt{2}m_{K}^{2}f_{K}-\frac{1}{\sqrt{2}} m_{\pi}^{2}f_{\pi}\right]
 -\sum_{i} g_{\zeta i}\rho_{i}^{s} = 0 ,
 \label{zi_eq}
\end{align}
\begin{align}
&k_{0}\chi^{2}\delta-4k_{1}\left( \sigma^{2}+\zeta^{2}+\delta^{2}\right)
\delta-2k_{2}\left( \delta^{3}+3\sigma^{2}\delta\right) +2k_{3}\chi\delta
\zeta \nonumber\\
 + &  \frac{2}{3} d \chi^4 \left( \frac{\delta}{\sigma^{2}-\delta^{2}}\right)
-\sum_{i} g_{\delta i}\tau_3\rho_{i}^{s} = 0 ,
\label{di_eq}
\end{align}
\begin{align}
&k_{0}\chi \left( \sigma^{2}+\zeta^{2}+\delta^{2}\right)-k_{3}
\left( \sigma^{2}-\delta^{2}\right)\zeta + \chi^{3}\left[1
+{\rm {ln}}\left( \frac{\chi^{4}}{\chi_{0}^{4}}\right)  \right]
+(4k_{4}-d)\chi^{3}
\nonumber\\
-&\frac{4}{3} d \chi^{3} {\rm {ln}} \Bigg ( \bigg (\frac{\left( \sigma^{2}
-\delta^{2}\right) \zeta}{\sigma_{0}^{2}\zeta_{0}} \bigg )
\bigg (\frac{\chi}{\chi_0}\bigg)^3 \Bigg )+
\frac{2\chi}{\chi_{0}^{2}}\left[ m_{\pi}^{2}
f_{\pi}\sigma +\left(\sqrt{2}m_{K}^{2}f_{K}-\frac{1}{\sqrt{2}}
m_{\pi}^{2}f_{\pi} \right) \zeta\right] \nonumber\\
-& \frac{\chi}{{\chi^2}_0}(m_{\omega}^{2} \omega^2+m_{\rho}^{2}\rho^2)  = 0 ,
\label{chi_eq}
\end{align}
\begin{align}
\Bigg(\frac{\chi}{\chi_{0}}\Bigg)^2m_{\omega}^2 \omega +4g_{4}(\omega^3 + 3\omega\rho^2)-\sum_{i}{g_{\omega i}\rho_{i}}=0,
\label{omega_eq}
\end{align}
\begin{align}
\Bigg(\frac{\chi}{\chi_{0}}\Bigg)^2m_{\rho}^2\rho +4g_{4}(\rho^3 + 3\omega^2\rho)-\sum_{i}{g_{\rho i}\tau_{3}\rho_{i}}=0.
\label{rho_eq}
\end{align}

The effective nucleon masses $m_i^*$ can be generated with the help of scalar fields except for a small explicit mass term and this effective mass will decrease with the increasing baryonic density due to the partial restoration of chiral symmetry  \cite{Schramm2008} and is written as 
\begin{align}
m_{i}^{*}=g_{\sigma i}\sigma +g_{\zeta i}\zeta +g_{\delta i}\tau_{3}\delta +\delta m,
\label{eq7}
\end{align}
where $\delta m$ is a constant term added in order to get the vacuum mass of the nucleon. Moreover, $\rho_{i}$ and $\rho_{i}^{s}$ are vector and scalar densities of $i^{th}$ nucleon \cite{Papazoglou}.
\Cref{si_eq} to \cref{rho_eq}  are solved simultaneously along with the conditions of charge neutrality and beta-equilibrium to obtain density dependent values of various scalar and vector fields relevant for medium of neutron stars.
\par
Now, we discuss the strategy to evaluate the scalar quark and gluon condensates,  in terms of above obtained scalar fields, to be used later in QCD sum rule calculations for $\rho$ mesons. In chiral effective model, the explicit symmetry breaking is used to calculate the scalar quark condensates while the scalar gluon condensates can be obtained through the broken scale invariance property of QCD \cite{kumar2015}. The scalar quark condensates are related to explicit symmetry breaking through relation  \cite{Kumar2010}
\begin{equation}
\sum_{i} m_{i}\langle \bar{q}_{i}q_{i}\rangle=-\mathcal{L}_{SB},
\label{eq8}
\end{equation}
where $\mathcal{L}_{SB}$ is given by \cref{exp_lterm}.
%The explicit symmetry breaking term of the chiral effective Lagrangian is given by
%\begin{equation}
%\mathcal{ L} _{SB}= -\left( \frac {\chi}{\chi_{0}}\right)^{2} 
%\left( \frac{1}{2} m_{\pi}^{2} 
%f_{\pi} \left( \sigma + \delta \right) +
%\frac{1}{2} m_{\pi}^{2} 
%f_{\pi} \left( \sigma - \delta \right)
% + \big( \sqrt {2} m_{K}^{2}f_{K} - \frac {1}{\sqrt {2}} 
%m_{\pi}^{2} f_{\pi} \big) \zeta \right). 
%\end{equation}
From this one can identify following expressions for light quark condensates \cite{kumar2015}:
\begin{equation}
m_{u}\langle \bar{u}u\rangle =\frac{1}{2}m_{\pi}^2 f_{\pi}(\sigma + \delta)\Bigg(\frac{\chi}{\chi_{0}}\Bigg)^{2},
\label{eq_u_cond}
\end{equation}
\begin{equation}
m_{d}\langle \bar{d}d\rangle =\frac{1}{2}m_{\pi}^2 f_{\pi}(\sigma - \delta)\Bigg(\frac{\chi}{\chi_{0}}\Bigg)^{2},
\label{eq_d_cond}
\end{equation}
\begin{equation}
m_{s}\langle \bar{s}s\rangle =\Big(\sqrt{2}m_{K}^2f_{K}-\frac{1}{\sqrt2}m_{\pi}^2f_{\pi}\Big)\Bigg(\frac{\chi}{\chi_{0}}\Bigg)^{2}\zeta.
\end{equation}
The energy-momentum tensor $\theta_{\mu \nu}$ in terms of dilaton field in effective chiral model is given as 
\begin{equation}
\theta_{\mu \nu}=(\partial _\mu \chi) 
\Bigg (\frac {\partial {\mathcal{L_{\chi}}}}
{\partial (\partial ^\nu \chi)}\Bigg )
- g_{\mu \nu} \mathcal{L_{\chi}},
\label{eq9}
\end{equation}
where $\mathcal{L_{\chi}}$ is the Lagrangian due to the dilaton field and is given by  \cite{Rajesh}
\begin{equation}
\mathcal{L_{\chi}}= \frac {1}{2} (\partial _\mu \chi)(\partial ^\mu \chi)- k_4 \chi^4 - \frac{1}{4} \chi^{4} {\rm {ln}} 
\Bigg ( \frac{\chi^{4}} {\chi_{0}^{4}} \Bigg )
+ \frac {d}{3} \chi^{4} {\rm {ln}} \Bigg (\bigg( \frac {\left( \sigma^{2} 
- \delta^{2}\right) \zeta }{\sigma_{0}^{2} \zeta_{0}} \bigg) 
\bigg (\frac {\chi}{\chi_0}\bigg)^3 \Bigg ).
\label{eq10}
\end{equation} 
The scalar gluon condensate is equal to the trace of energy-momentum tensor which is non-vanishing in the present case. This property of non-vanishing trace of $\theta_{\mu \nu}$ is known as the trace anomaly which arises due to the broken scale invariance property of QCD \cite{kumar2015}. Using the broken scale invariance Lagrangian, the trace of energy-momentum tensor in the limit of vanishing quark masses is given by 
\begin{equation}
\theta_{\mu}^{\mu} = (\partial _\mu \chi) \Bigg (\frac {\partial {\mathcal{L_{\chi}}}}
{\partial (\partial _\mu \chi)}\Bigg ) -4 {\mathcal{L_{\chi}}}=-(1-d)\chi^{4}. 
\label{eq11}
\end{equation}
Comparing the trace of of energy-momentum tensor in QCD with that calculated in effective chiral model, the gluon condensate found out to be \cite{Mishra,Kumar}
\begin{equation}
\Bigg\langle \frac{\alpha_s}{\pi}G_{\mu \nu}^a G^{a \mu \nu}\Bigg\rangle=\frac{8}{9}(1-d)\chi^4.
\label{eq12}
\end{equation}
The scalar gluon condensate can be modified by incorporating the finite mass of quarks and its expression becomes  \cite{Kumar}
\begin{equation}
\Bigg\langle \frac{\alpha_s}{\pi}G_{\mu \nu}^a G^{a \mu \nu}\Bigg\rangle=\frac{8}{9}\Bigg[(1-d)\chi^4 + \Big(m_{\pi}^2f_{\pi}\sigma+\Big(\sqrt{2}m_{K}^2f_{K}-\frac{1}{\sqrt2}m_{\pi}^2f_{\pi}\Big)\zeta\Big)\Bigg(\frac{\chi}{\chi_{0}}\Bigg)^{2}\Bigg].
\label{eq13}
\end{equation}

\subsection{QCD sum rules} 
\label{subsec:2.2}
\quad In this subsection, we will discuss briefly the QCD sum rules used in the present work to calculate the in-medium masses of vector $\rho$ mesons in charge neutral matter \cite{Zschocke2002,Lee,Mishra,Kwon}. To obtain the QCD sum rules for $\rho$ mesons, we start with the retarded two-point correlation function defined in terms of vector current $j^{\rho}_{\mu}$ corresponding to
$\rho$ mesons at finite baryonic density as \cite{Zschocke2002}:
\begin{equation}
 \Pi_{\mu \nu}^{\rho R}(q ;\rho_{B})=i \int d^4x e^{iqx} \langle\mathcal{R} j^{\rho}_{\mu}(x)j^{\rho}_{\nu}(0) \rangle_{\rho_{B}},
 \label{eq_cor1}
\end{equation} 
where $q = (q^0,\textbf{q})$, $\langle ... \rangle_{\rho_{B}}$ is the grand canonical ensemble average, $\rho_{B}$ is the baryonic density, $\mathcal{R} j^{\rho}_{\mu}(x)j^{\rho}_{\nu}(0)\equiv \Theta (x^{0})[j^{\rho}_{\mu}(x),j^{\rho}_{\nu}(0)]$ with  $j^{\rho}_{\mu}$=$\frac{1}{2}$($\bar{u}\gamma_{\mu}u$ - $\bar{d}\gamma_{\mu}d$) and $\Theta(x^{0})$ is the step function.\\ In the limit $\textbf{q}\rightarrow 0 $  the longitudinal and transverse invariant relevant for the medium will merge into  single one  given by \cite{Lee}
\begin{align}
\Pi^{\rho R}=\Pi_{\mu \mu}^{\rho R}/(-3q^{2})|_{\textbf{q}\rightarrow 0}.
\label{eq14}
\end{align}
Moreover, since the complex $q^{0} (=\omega)$ plane is analytic in the upper half plane,  following dispersion relation  can be written for the correlation function in the medium \cite{Hatsuda,Zschocke2002}   
\begin{align}
\text{Re}\Pi^{\rho R}(\omega^{2})=  P\int_{0}^{\infty} du^{2} \frac{R_{h}^{\rho}}{u^{2}-\omega^{2}} + \text{subtractions}.
\label{eq_cor2}
\end{align}
In above, on right side (known as phenomenological side) $R_{h}^{\rho}=\frac{1}{\pi}$Im$\Pi^{\rho R}(u)$ (where $h$ stands for hadrons) is the spectral density. In the free space, the spectral density is generally written as sum of resonance pole term and continuum term, whereas in the charge neutral matter in which we are interested in the present work, additional contribution, known as Landau damping contribution, from the scattering of mesons with nucleons will need to consider \cite{Zschocke2002}. 
Considering these three contributions, we can write
\begin{align}
12 \pi \text{Im} \Pi^{\rho R}(u) = \delta(u^2) R_{\text{sc}} + F_{\rho}^* \delta(u^2 - m_\rho^{*2}) + c_0 \theta (u^2 - s_{0}^{*\rho}),
\label{eq15}
\end{align}
where $R_{\text{sc}}$  = $3 \pi^2 \frac{\rho_B}{m_N}$ is Landau damping contribution \cite{Hatsuda}. 
Also,  $m_{\rho}^{*}$ is the in-medium mass of $\rho$ meson, $F_{\rho}^{*}$ is the pole residue and $s_{0}^{* \rho}$ is the continuum threshold which separates the resonance part and the perturbative part of spectral density.
The parameters $m_\rho^{*}$, $F_{\rho}^*$ and 
$s_{0}^{*\rho}$ are known as phenomenological parameters.
The subtraction term in \cref{eq_cor2} accounts for the convergence of the integral. If $\text{Im}\Pi^{\rho R}(u)$ in the dispersion integral diverges then subtraction term exists otherwise it will be zero \cite{AV}. Assuming the integral to be converging in the present investigation, we used subtraction term to be zero \cite{Zschocke2002}. 
\par
For the large values of $Q^2$ (i.e. $Q^{2}\equiv-q_{0}^{2}>0$), we perform the operator product expansion (OPE) of
the product of vector current appearing in \cref{eq_cor1}. Thus, for the left hand side of \cref{eq_cor2}, we have 
\begin{align}
\text{Re}\Pi^{\rho R}(Q^{2})=-c_{0}\text{ln}Q^{2}+\sum_{n=1}^{\infty}\frac{c_{n}}{Q^{2n}},
\end{align}
where $c_{n}$ includes the Wilson coefficients \cite{HL} and include the contribution in terms of non-zero quark and gluon condensates.  On comparing the phenomenological (right side) and OPE side (left side) of \cref{eq_cor2} in asymptotic limit, we obtain following set of finite energy sum rule equations \cite{Zschocke2002, Mishra, Kraniskov}:
\begin{equation}
F_{\rho}^{*}=(c_{0} s_{0}^{*\rho}+c_{1} )- R_{\text{sc}},
\label{eq_fser1} 
\end{equation}
\begin{equation}
F_{\rho}^{*}m_{\rho}^{*2}=\frac{(s_{0}^{*\rho})^2c_{0}}{2}-c_{2}^{*},
\label{eq_fser2}
\end{equation}
\quad and
\begin{equation}
F_{\rho}^{*}m_{\rho}^{*4}=\frac{(s_{0}^{*\rho})^3c_{0}}{3}+c_{3}^{*}.
\label{eq_fser3}
\end{equation}
  The Wilson coefficients $c_n$ appearing in the above given sum rule equations upto $n=3$ are given by 
\begin{equation}
c_{0}=\frac{3}{2}\Bigg[1+\frac{\alpha_{s}(Q^2)}{\pi}\Bigg] ,\quad\quad c_{1}=\frac{-9}{2}(m_{u}^2+m_{d}^2), \\
\label{sum1}
\end{equation}
\begin{equation}
c_{2}^{*} = \frac{\pi^2}{2}\Big\langle\frac{\alpha_{s}}{\pi} G_{\mu \nu}^a G^{a\mu \nu}\Big\rangle+6\pi^2\langle m_{u} \bar{u}u + m_{d}\bar{d}d \rangle,
\label{sum2}
\end{equation}
\begin{equation}
c_{3}^{*} = -\alpha_{s}\pi^3\frac{224}{27}k_{q}(\langle\bar{u}u\rangle^2+\langle\bar{d}d\rangle^2),
\label{sum3}
\end{equation}
where $\alpha_{s}$ is the coupling constant, $m_{u}$ and $m_{d}$ are the masses of up and down quarks with respective values of 4 and 7 MeV and $k_{q}$ is the factorization parameter.
In the above discussion, the value of $c_{3}^{*}$ represent the 4-quark condensates and in calculating this we need the factorization parameter $k_{q}$ which is calculated first by assuming the vacuum case only \cite{Mishra}. In our calculations, we have used a value of 2.0850 for $k_{q}$.

\section{Results and Discussion}
\label{sec:3}
\begin{table}
\begin{tabular}{|c|c|c|c|c|}
\hline 
$g_{\sigma N}$  & $g_{\zeta N }$  &  $g_{\delta N }$  &
$g_{\omega N}$ & $g_{\rho N}$ \\

\hline 
-9.83 & 1.22 & -2.34 & 12.13 & 4.03  \\

\hline
$k_0$ & $k_1$ & $k_2$ & $k_3$ & $k_4$  \\ 
\hline 
2.37 & 1.40 & -5.55 & -2.65 & -0.23  \\ 

\hline 
$m_\pi $(MeV) &$ m_K$ (MeV) &$ f_\pi$ (MeV) & $f_K$(MeV) & $g_4$ \\ 
\hline 
139 & 498 & 93.3 & 122.143 & 38.9  \\ 
\hline

$\sigma_0$ (MeV) & $\zeta_0$ (MeV) & $\chi_0$ (MeV) & $d$ & $\rho_0$ ($\text{fm}^{-3}$)  \\ 
\hline 
-93.3 & -106.8 & 401.91 & 0.06 & 0.15  \\ 

\hline
$m_{\omega} $(MeV) &$ m_{\rho}$(MeV) &$m_{N}$(MeV) &$\delta m$(MeV) &$k_{q}$\\
\hline
783 & 770 & 939 & 150 & 2.0850 \\
\hline

\end{tabular}
\caption{Values of various parameters.}
\label{table1}
\end{table} 

\begin{figure}
\includegraphics[width=12cm,height=10cm]{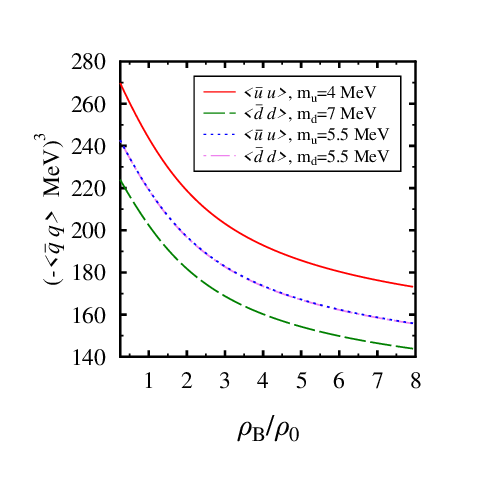}
\caption{The scalar quark condensates are plotted against the increasing baryonic density in neutron star matter.}
\label{fig1}
\end{figure}

\begin{figure}
\includegraphics[width=12cm,height=10cm]{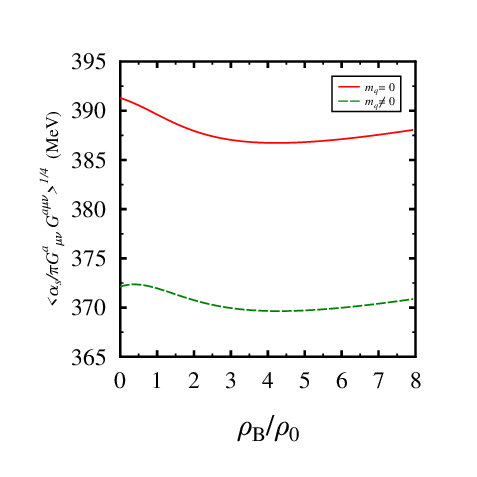}
\caption{The gluon condensates are plotted against the increasing baryonic density in neutron star matter.}
\label{fig2}
\end{figure}

%\begin{figure}
%\includegraphics[width=9cm,height=6cm]{c_2.eps}
%\caption{(Color online)The coefficient $c_2$ is plotted against the increasing baryonic density in neutron star matter.}
%\label{fig6}
%\end{figure}

\begin{figure}
\includegraphics[width=12cm,height=10cm]{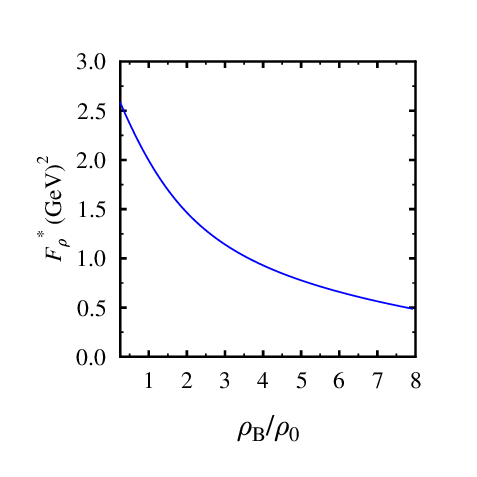}
\caption{Variation of pole residue $F_{\rho}^{*}$ with the increasing baryonic density in neutron star matter is shown
in above figure.}
\label{fig3}
\end{figure}

\begin{figure}
\includegraphics[width=12cm,height=10cm]{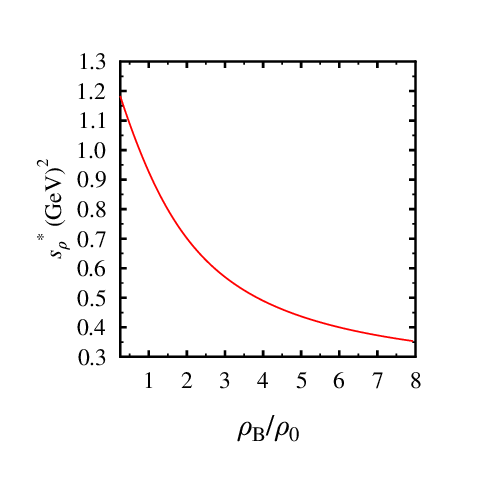}
\caption{Variation of continuum threshold parameter $s_{\rho}^{*}$ with the increasing baryonic density in neutron star matter.}
\label{fig4}
\end{figure}

\begin{figure}
\includegraphics[width=12cm,height=10cm]{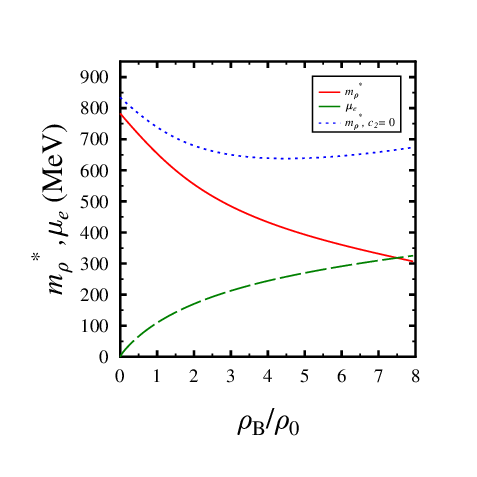}
\caption{The in-medium mass of $\rho$ meson, $m_{\rho}^{*}$ and the electron chemical potential, $\mu_e$  are plotted against the increasing baryonic density in neutron star matter.}
\label{fig5}
\end{figure}

%\begin{figure}
%\includegraphics[width=12cm,height=10cm]{mass_difference_test.eps}
%\caption{(Color online)Dependence of in-medium mass of $\rho$ meson on Wilson coefficients }
%\label{fig6}
%\end{figure}

\begin{table}
\begin{tabular}{|c|c|c|c|c|}
\hline
$c_{2\downarrow}^{*},\quad m_{\rho}^{*}\rightarrow $ & $\rho_{B}=\rho_{0}$ & $\rho_{B}=2\rho_{0}$ & $\rho_{B}=4\rho_{0}$ \\ 
\hline 
$c_{2}^{*}$=0 & 738.7 (MeV), (11.5$\%$) & 679.3 (MeV), (18.7$\%$) & 638.9 (MeV), (23.4$\%$) \\ 
\hline
$c_{2}^{*}\ne0$ & 653.9 (MeV),(15.8$\%$) & 554.8 (MeV), (28.5$\%$) & 433.1 (MeV), (44.2$\%$)  \\
\hline
\end{tabular}
\caption{Variation of mass with $c_{2}^{*}$}
\label{table2}
\end{table}

Here, we will present the results of our present investigation of $\rho$ mesons and possibility of its condensation in neutron star medium. As discussed earlier, the quark and gluon condensates to be used in QCD sum rules to obtain the in-medium masses of $\rho$ mesons are calculated using chiral SU(3) model. In \cref{fig1,fig2} we have shown the variation of scalar quark condensates and gluon condensates, respectively as a function of baryonic density $\rho_B$ of charge neutral matter.\\
As can be seen from \cref{fig1}, as the density of matter is increased the magnitude of quark condensate decrease. At $\rho_{B}=0, \rho_{0}$ and $4\rho_{0}$ the values of $\langle \bar{u}u\rangle$ are observed to be (-279.6 MeV)$^{3}$, (-243.6 MeV)$^{3}$ and (-192.9 MeV)$^3$, respectively. For $\langle \bar{d}d\rangle$ above values changes to (-232.1 MeV)$^{3}$, (-202.3 MeV)$^{3}$ and (-160.2 MeV)$^{3}$, respectively. The drop in the magnitude of quark condensates as a function of density is consistent with the expectation of chiral symmetry restoration at high baryonic density. As can be seen from \cref{eq_u_cond,eq_d_cond}, $\langle \bar{u}u\rangle$ (or $\langle \bar{d}d\rangle$) condensate is directly proportional to the scalar field $\sigma$ (note that the magnitude of $\delta$ field is very small as compared to $\sigma$ field and also, the scalar dilaton field $\chi$ vary very little as function of density) whose in-medium value is calculated by solving coupled equations of scalar and vector fields in charge neutral matter, as was discussed earlier also. The magnitude of  scalar field $\sigma$ is observed to decrease with increase in the density of medium and hence causes decrease in the magnitude of scalar quark condensates. 
 The difference in the vacuum values ($\delta$ field will be zero at $\rho_B$ =0) of $\langle \bar{u}u\rangle$ and $\langle \bar{d}d\rangle$ quark condensates quoted above is because 
 we considered different current quark masses of $u$ and $d$ quarks. Having considered average value of two masses, $m_{q}=m_{u}=m_{d}=5.5$ MeV, vacuum value of these quark condensates will coincide whereas in the medium they still have different values  because of iso-vector field $\delta$ appearing with opposite sign in \cref{eq_u_cond,eq_d_cond}.
%  For this case, $\langle \bar{q}q\rangle$ has magnitude greater than the $\langle \bar{d}d\rangle$ but less than $\langle \bar{u}u\rangle$ as also evident from the values at $\rho_{B}=0, \rho_{0}$ and 4$\rho_{0}$ which are (-251.5 MeV)$^{3}$, (-219.1 MeV)$^{3}$ and (-173.4 MeV)$^{3}$ respectively.
  In \cite{Bali,SS,Hong,DH,Sho}, the density dependence of quark condensates is evaluated for nuclear medium relevant for heavy-ion collisions. Calculations of \cite{Walecka,Serot} within linear Walecka model and of \cite{NK,DELFINO}  using Born potential in Dirac-Brueckner approach show that quark condensates first decreases with density and then start increasing at higher baryonic density. 
%  The difference in the values of the $\langle \bar{u}u\rangle$ and $\langle \bar{d}d\rangle$ condensates can also be foreseen from the corresponding relations Eqs. (18) and (19) as for a charge neutral matter the value of $\delta$ field is not zero and hence the two condensates differs.

\par As can be seen from \cref{fig2}, the value of scalar gluon condensate $\langle\frac{\alpha_{s}}{\pi}G_{\mu\nu}^a G^{a\mu\nu}\rangle$ also shows a decrease with respect to vacuum value. We calculated the gluon condensates in charge neutral medium, with and without quark mass term. 
%From Fig.(\ref{fig2}) we can see that the magnitude of $\langle\frac{\alpha_{s}}{\pi}G^{\mu\nu}G_{\mu\nu}\rangle$ shows an irregular behaviour with the increasing baryonic density. 
We found that the value of gluon condensate, without quark mass term, at density $\rho_{B}=0, \rho_{0}$, $4\rho_{0}$ and $6\rho_{0}$ is (391.3 MeV)$^{4}$, (389.6 MeV)$^{4}$, (386.7 MeV)$^{4}$ and (387.1 MeV)$^{4}$, respectively. The corresponding values of $\langle\frac{\alpha_{s}}{\pi}G_{\mu\nu}^a G^{a\mu\nu}\rangle$, considering finite quark mass term, are (372.9 MeV)$^{4}$, (372.5 MeV)$^{4}$, (369.9 MeV)$^{4}$ and (370.2 MeV)$^{4}$, respectively. The dilaton field $\chi$ varies very little as a function of density of medium and this behaviour is further reflected in
the density dependence of scalar gluon condensates which is also in agreement with the calculations in refs. \cite{Thomas, AK}.
%In earlier calcu the linear density approximation dependence on condensates as were in earlier studies \cite{Hatsuda,Weise} which results in more decrease on the values of these condensates. From Fig. (\ref{fig2}) we can see that the value of gluon condensates decreases in the low density regime but it will show an increasing behavior at higher density which is evident from the given values at different density. 
We should notice here that the variation of quark condensate is more in comparison to the gluon condensate. This can also be concluded from the drop in the two condensates at $\rho_{B}$=$\rho_{0}$, quark condensate drops to $\sim$13\% while gluon condensate to only $\sim$0.4\% of respective vacuum values.

In the following now, we will discuss the modification of the masses of $\rho$ mesons in charge neutral matter using the above discussed medium modified scalar quark and gluon condensates. Solving coupled finite energy sum rules given by \cref{eq_fser1,eq_fser2,eq_fser3}, we obtain the density dependence of pole residue $F_{\rho}^{*}$, continuum threshold parameter $s_{\rho}^{*}$ and in-medium vector $\rho$ meson  mass $m_{\rho}^{*}$ plotted in \cref{fig5,fig3,fig4}.
As can be seen from \cref{fig3,fig4}, there is a continuous decrease in the value of  $F_{\rho}^{*}$ and $s_{\rho}^{*}$ as a function of baryonic density. At density, $\rho_{B}$= 0, $\rho_{0}$ and 4$\rho_{0}$ the values of $F_{\rho}^{*}$ ($s_{\rho}^{*}$) are observed to be 2.817 (1.285), 1.994 (0.926) and 0.929 (0.489) GeV$^{2}$, respectively. Due to more scattering of nucleons at higher density the effect of Landau damping term becomes significant and thus the value of $F_{\rho}^{*}$ will decrease more at higher density \cite{Hatsuda}. 
%The value of $s_{\rho}^{*}$ which separates the resonance and continuum part is also plotted as a function of baryonic density in \cref{fig5} \cite{Hatsuda,Weise}. As the density increases, the scattering between increases and the separation between the resonance and the continuum part also decreases. 

\par The dependence of the in-medium masses of $\rho$ mesons on the condensates is introduced through the Wilson coefficients, $c_{2}^{*}$ and $c_{3}^{*}$ as given by \cref{sum2,sum3}. As also discussed in earlier calculations of ref. \cite{Hatsuda,Mishra}, the coefficient $c_{2}^{*}$ in \cref{sum2} can be written in terms of gluon condensates and the twist-two quark condensates, while the coefficient $c_{3}^{*}$ of \cref{sum3} can be written in terms of twist-four quark condensates. It can be inferred from the \cref{table2,fig5} that the mass of $\rho$ meson depends largely on $c_{2}^{*}$ at higher density. The drop in the mass of $\rho$ meson in the absence of $c_{2}^{*}$ at $\rho_{B}=\rho_{0}$, $2\rho_{0}$ and $4\rho_{0}$ is found to be 11.5$\%$, 18.7$\%$ and 23.4$\%$ of the vacuum value. On the other hand, in presence of $c_{2}^{*}$ the drop is more at higher density which is about 45$\%$ of the vacuum value, near the density of $4\rho_{0}$. In absence of $c_{2}^{*}$, the dependence is solely on the twist-four quark condensates which contributes positively to the mass, while in contrary to this, the presence of the $c_{2}^{*}$ at higher density will contribute more negatively through the twist-two quark condensates.

\par
From \cref{fig5}, we observe that the mass of $\rho$ meson decrease with increase in the density of the medium.
%  The possibility of $\rho$ meson condensation in a neutron star is also investigated and it is found that beyond a certain density the $\rho$ meson condensation is possible in neutron star (Fig.5). 
  At $\rho_{B}=0$, $\rho_{0}$ and 4$\rho_{0}$, it is found to have a value of 776.8, 653.9 and 433.1 MeV, respectively. The drop in the mass of $\rho$ mesons with increasing density is consistent with earlier QCD sum rule calculations \cite{Hatsuda,Koike,Zschocke2002}.
%  , Koike and Hayasizaki \cite{Koike},Zschocke and group \cite{Zschocke2002}, Asakawa and Ko \cite{Zschocke2002} in the linear density approximation.
In \cref{fig5}, we also plotted the chemical potential of electron as a function of density. 
On one hand, there is a drop in the in-medium mass of $\rho$ meson with the increasing baryonic density while on the other hand, the chemical potential of the electron increases with the increased density. The baryonic decay will give rise to the presence of electrons and these released electrons will require a larger amount of energy to exist at higher densities. Thus, at such high baryonic densities, it is not energetically favorable for electrons to be present and also in parallel to this effect, the drop in the mass of $\rho$ meson is appreciable which could make them a better contestant to exist in place of electrons. Thus, when the in-medium mass of $\rho$ mesons becomes comparable to the effective electron chemical potential these electrons may get replaced by the negatively charged $\rho$ mesons. As can be seen from \cref{fig5}, at baryonic density $\rho_{B}=7.3\rho_{0}$, the condition $\omega_{\rho}=\mu_{e^-}$ is satisfied and hence possibility of $\rho$ meson condensation arise at this density. Furthermore, the possibility of $\rho$ meson to be found at the core of a neutron star is also supported by few other works which includes \cite{EE,Mallick2015}. In ref. \cite{Mallick2015}, a more simpler approach is used to calculate the in-medium masses of $\rho$ mesons by considering the linear dependence on $\sigma$ field.

\par
The presence of strong magnetic field at the interior core of the neutron star may also effect the threshold density at which the $\rho$ meson condensation takes place \cite{Prakash, Cardall, Chakrabarty, Zhang, Mao, Wei, Iwamoto, Khalilov, Yue, Rabhi,S,Lattimer2}. 
 The observations of pulsars and gamma repeaters clearly indicate the presence of a strong magnetic field in neutron stars. It is estimated to be around $10^{14}$-$10^{15}$G at the surface \cite{Thompson} which becomes even more intense at the interior core. According to the virial theorem, it is approximated to be about $10^{18}$-$10^{19}$G in the interior core of the star \cite{Lai, Zhang}. The dense quark matter at the center is assumed to be the reason for such a high magnetic field. 
% Many theoretical researchers have also worked on to study the effect of magnetic field on the highly dense matter and neutron star \cite{Prakash, Cardall, Chakrabarty, Zhang, Mao, Wei, Iwamoto, Khalilov, Yue, Rabhi}.
  It was observed that the composition of neutron star could change efficaciously in the presence of such a high magnetic field \cite{Yue}. As discussed earlier, in various papers the effect of the Landau quantization due to such a high magnetic field on anomalous magnetic moments of nucleons could affect the EoS of the neutron star \cite{Prakash,S,Lattimer2}. In case of magnetized nuclear matter it was observed that the drop in the medium modified mass of $\rho$ meson is a bit low as compared to the non-magnetized neutron matter \cite{AK} which may slightly diminish the possibility of $\rho$ meson condensation in neutron star also. Furthermore, calculations of refs. \cite{Yuan,Prakash,Yue} have also shown the sharp decrease in the masses of nucleons in presence of magnetic field. As a result, the Fermi momenta of protons will fall and proton fraction will increase and to maintain the beta equilibrium condition electron chemical potential also which results in increase of electron density. Moreover, calculations in QMC model, Modified Quark-Meson Coupling (MQMC) and RMF model \cite{Shen,Ryu,Dey} 
    show that the threshold density for kaon and antikaon condensation  shift to higher values in presence of magnetic field which results in the large fraction of electrons at higher densities. This causes more stiffness in the EoS and increase in the mass of neutron star. Similar kind of consequences may arise due to $\rho$ meson condensation in neutron stars when finite magnetic field effects will be considered in future detailed calculations.  
%  
%  Moreover, calculations in QMC model and RMF model in refs. \cite{Shen} and \cite{Dey} respectively shows that the threshold density for kaon and antikaon condensation also shift to higher values in presence of magnetic field which results in the large fraction of electrons at higher densities  and also in ref. \cite{Ryu} using Modified Quark-Meson Coupling (MQMC) it is shown that at higher magnetic field the kaon condensation gets suppressed by the presence of electron. At higher densities due to the presence of kaons the EoS becomes stiff which results in a larger mass of neutron star. This will also suggest a hindrance to the possibility of $\rho$ meson condensation in neutron star matter. The presence of $\rho$ mesons may also directly influence the equation of state just as the presence of kaons.
\par
Contrary to the neutron stars, proto-neutron star which is the initial stage of the neutron star just after the supernova explosion include neutrinos as an integral part of it. Calculations of refs. \cite{Pons,Schramm2008} shows that as a consequence of increased number of neutrinos the electron density should also increase and therefore demanding the charge neutrality condition proton density should also increase. The increasing proton density will decrease the isospin asymmetry in the star which will soften the EoS. Due to the large population of electron at lower density, it may also diminish the possibility of presence of $\rho$ mesons.

%Moreover, since we are aware of the fact that quarks are the charged particles and they can get affected by the magnetic field easily. The most important thing needed to study the neutron star properties is its EoS \cite{Lattimer2}.
%\section{Summary}
%\label{sec:4}
In brief, we have investigated the possibility of $\rho$ meson condensation in neutron star using a combined approach of chiral SU(3) model and QCD sum rules. In this, chiral SU(3) model is employed to calculate the scalar quark and gluon condensates through scalar and vector fields in charge neutral matter. The condensates are used in QCD sum rules through the Wilson coefficients, to determine the in-medium masses of $\rho$ mesons. The effect of scattering between the nucleons is incorporated through the Landau damping term, $R_{sc}$. An appreciable decrease in the mass of $\rho$ meson is observed and alongside to this, the electron chemical potential as calculated by solving the chiral model in neutron star shows an increasing behaviour with increasing baryonic density. The threshold density at which condition of $\rho$ meson condensation  is satisfied is found to be 7.3$\rho_{0}$.
% It is further investigated that in absence of magnetic field, the in-medium mass of $\rho$ meson drops significantly and at density, $\rho_{B}$=7.3$\rho_{0}$ the condition for $\rho$ meson condensation ($\omega_{\rho}=\mu_{e^{-}}$) is established.
% Moreover, the investigation of Wilson coefficients shows the dependence of in-medium mass on the twist-two quark condensate and twist-four condensate. At higher density, the drop is high in the presence of twist-two quark condensate which is in contrast to the case of zero twist-two quark condensate. Furthermore, the density dependent condensates are also discussed within the neutron star matter in this investigation. It is observed that the scalar quark condensate directly affect the in-medium masses of $\rho$ mesons while the variation in the gluon condensate is not much effective solely.
%We find that due to enough drop in the in-medium of masses of $\rho$ mesons and the corresponding increase in chemical potential of the electron can give rise to $\rho$ meson condensation at about 7$\rho_{0}$ while on comparing the result of present investigation with the magnetized nuclear matter we can deduce that the possibility of $\rho$ meson condensation in neutron star is less probable.

\end{document}